%% file: main.tex
\documentclass[conference]{IEEEtran}

\usepackage{cite}
\usepackage{amsmath}
\usepackage{algorithmic}
\usepackage{array}
\usepackage[caption=false,font=footnotesize]{subfig}
\usepackage{url}
\usepackage{graphicx}
\usepackage{subfiles}
\usepackage{amsthm, amssymb}
\usepackage{textcomp}
\usepackage{xcolor}
\usepackage[utf8]{inputenc}
\usepackage[T1]{fontenc}
\usepackage{balance}
\usepackage{xspace}
\usepackage{multirow}
\usepackage{float}
\usepackage{soul}
\usepackage{paralist}

\def\BibTeX{{\rm B\kern-.05em{\sc i\kern-.025em b}\kern-.08em
    T\kern-.1667em\lower.7ex\hbox{E}\kern-.125emX}}

\theoremstyle{definition}
\newtheorem{definition}{Definition}
\begin{document}

\title{Towards Evidence-based Testability Measurements}

\author{\IEEEauthorblockN{Luca Guglielmo, Andrea Riboni, Giovanni Denaro}
\IEEEauthorblockA{Department of Computer Science, Systems and and Communication (DISCo)\\University of Milano-Bicocca\\
Viale Sarca, 336, 20125 Milan (MI), Italy\\
Email: l.guglielmo@campus.unimib.it, andrea.riboni@unimib.it, giovanni.denaro@unimib.it}}

\maketitle

\begin{abstract}
Evaluating Software testability can assist software managers in optimizing testing budgets and identifying  opportunities for refactoring.
In this paper, we abandon the traditional approach of pursuing testability measurements based on the correlation between software metrics and test characteristics observed on past projects, e.g., the size, the organization or the code coverage of the test cases. We propose a radically new approach that exploits automatic test generation and mutation analysis to quantify the amount of evidence about the relative hardness of identifying effective test cases. We introduce two novel evidence-based testability metrics, describe a prototype to compute them, and discuss initial findings 
on whether our measurements can reflect actual testability issues.
\end{abstract}


\section{Introduction}\label{intro}
\subfile{section/intro}

\section{Evidence-based Testability Measurement}\label{sec:approach}
\subfile{section/approach}

\section{Initial Findings}\label{sec:experiment}
\subfile{section/exp}

\section{Conclusion and Future Work}\label{sec:conclusions}
\subfile{section/conclusion}

\bibliographystyle{IEEEtran}
\bibliography{bibliography}

\end{document}

%% file: section/intro.tex
Accomplishing effective software testing has a key role in producing high quality software at the state of the practice. Thus, being able to measure the \emph{testability} of the software artifacts under test, i.e., \emph{the degree to which the design of the artifacts supports or hardens their own testing}\cite{freedman1991testability,voas1995software,InternationalStandardOrganization(ISO).2001,IEEE1990}, can be a crucial information for managers. For example, the early availability of testability measures can enable informed decisions on optimising the testing budget, or pinpoint components that shall undergo refactoring before testing.

So far, the research on measuring software testability focused on  either i)~exploiting the correlation between static software metrics and testing effort or quality, or ii)~estimating the likelihood of detecting faults, in particular with reference to the \textit{execute-infect-propagate} (PIE) model of fault sensitivity. Static software metrics capture the static structure of software artifacts, e.g., the amount of lines of code, the McCabe's cyclomatic complexity, or the Chidamber and  Kemerer's class-level metrics~\cite{mccabe1976complexity,chidamber1994metrics}. Indeed, empirical data collected out of many past software projects support the existence of correlation between static metrics and the number, the size, the complexity, the code coverage or the mutation scores of the test suites~\cite{garousi2019survey,bruntink2004predicting,bruntink2006empirical,singh2008predicting,toure2018predicting, alshahwan2009improving,da2017empirical,jalbert2012predicting,khoshgoftaar2000predicting,yu2016predicting,terragni2020measuring}.

The PIE model defines fault sensitivity as the combined probability of executing faulty locations, infecting the execution state and propagating the effects of the infection to some observable output~\cite{voas1992pie}. 
High fault sensitivity can be a proxy of high testability, and vice-versa~\cite{voas1992pie,voas1991predicting,tsai2009study,binder1994design}.

This paper makes the observation that the research done so far addresses testability only indirectly, either in terms of predictions about size, complexity and coverage of test cases, or by estimating fault sensitivity scores. However, we cannot take for granted that these indirect measurements capture testability to the full extent. Moreover, the measurement approaches depend on the characteristics of the available test cases (e.g., test cases from past projects) which can be easily affected by arbitrary decisions of the testers (e.g., decisions on designing few or many test cases, or aiming to  high code coverage or ignoring code coverage). Yet, with reference to the body of literature on the correlation with software metrics, we recall that correlation does not necessarily entails causation, and in fact several studies yield contrasting results on which metrics are best predictors of testability, and predictive models trained on the data of past projects rarely generalise with stable precision~\cite{garousi2019survey}. 

This paper introduces a radically new idea on how to pursue software testability measurements. We aim at directly sampling the relative easiness (or the hardness) of identifying test cases for revealing the potential faults in the software modules under test. The higher the evidence of hard-to-test faults in a module, the higher the evidence that the design of that  module is not facilitating the testing. On this basis we refer to our approach as \emph{evidence-based testability measurement}.

Drawing on this idea, we propose an approach that simulates faults based on mutation-based fault seeding~\cite{just2014major,andrews2005mutation}, computes test cases in the style of search-based test generation techniques~\cite{fraser2011evosuite}, and discriminates between easy-to-test or hard-to-test faults based on the results of the test generator. In particular, we classify the seeded faults as  easy-to-test when the test generator is able to synthesize fault-revealing test cases out-of-the-box. Conversely, we classify the seeded faults as hard-to-test, if the test generator reveals them only when assisted with artificial testability boosters, like programming interfaces that relax the encapsulation constraints of the tested programs, or ideal test oracles that can predicate on intermediate execution states.  

The paper is organized as follows. 
Section \ref{sec:approach} presents our approach in detail, and contextualizes it on the problem of measuring testability for object oriented classes. Section \ref{sec:experiment} reports initial findings from a qualitative study on the classes of the \textit{Closure Compiler} project. Section \ref{sec:conclusions} summarizes our conclusions and future research plans.

%% file: section/approach.tex
\graphicspath{ {./sources/} }

This section discusses the intuitions that underlie our approach, and formalizes those intuitions into a reference theory of \textit{evidence-based testability metrics}. Next, we define a measurement framework that instantiates the reference theory heuristically, and present a prototype implementation. 

\subsection{The Rationale of Evidence-Based Testability Measurements}\label{rationale}

Our intuition is that software testability could be \emph{directly} quantified if we might know in advance
\begin{inparaenum}[(i)]
\item which faults may exist the software modules under test,
\item which test cases the developers could use, and
\item some criterion 
to analyze the test cases and argue whether they are easy or hard to be identified.
\end{inparaenum}
In this (admittedly) idealistic scenario, we could quantify the relative testability of the software modules based on the portion of easy-to-test and hard-to-test faults in each module. 
Intuitively, the higher the portion of hard-to-test faults in a software module, the higher the likelihood that the module may come with testability issues, and vice-versa. 

Drawing on these intuitions, we propose to evaluate testability as the portion of faults for which there exists some evidence (i.e., at least a test case) that those faults can be revealed with easy test cases. This is why we refer out testability metrics as \emph{evidence-based}. 
We define the following notion of testability:

\begin{definition}{\bf Idealistic Testability}\label{def:testability}
Let $M$ be a software module of a program $P$, $F$ be the set of \emph{executable}\footnote{Non-executable faults are irrelevant for testing and testability.} faults in $P$,  $T$ be the set of possible test cases for $P$, and $Reveal: F \times T$ be the relation between  faults and  test cases that reveal them. 
Let also $F(M)\subseteq F$ denote the faults located in $M$.

Moreover, let  $Hard: T \rightarrow \{true, false\}$ be a criterion (a predicate) to decide whether a test case is or is not hard to be identified. Accordingly, let  $F_{hard}(M)$ be the set of faults in $M$ that are hard to identify, that is, \(F_{hard}(M) \equiv \{f\in F(M) | \forall{t}: Reveal(f, t) \Rightarrow Hard(t)\}\).\footnote{The set $F_{hard}(M)$ includes also the faults that, although being executable, cannot be revealed with any test case. Executable, non-revealable faults are arguably synthoms of testability issues.}

Then the testability of  module $M$ is quantified as
\[Testability(M) = 1 - \frac{|F_{hard}(M)|}{|F(M)|}.\]
\end{definition}

We can further adapt the above definition to discriminate between 
\begin{inparaenum}[(i)]
\item \emph{controllability} issues, i.e., testability problems that depend on the difficulty of identifying test cases that execute the faults, and
\item \emph{observability} issues, i.e., testability problems that depend on the difficulty of identifying test cases that ultimately reveal the faults by producing observable malfunctions.
\end{inparaenum}

\begin{definition}{\bf Idealistic Controllability}\label{def:contr:ideal}
Let $Exec: F \times T$ be the relation between faults and test cases that execute them, and  $Hard\_d: T \rightarrow \{true, false\}$ be a criterion to decide whether or not the driver part of a test case is hard to be identified. The driver is the part of a test case that sets proper inputs for the module under test, aiming to drive its execution in specific way. Correspondingly, let  $F_{hard\_d}(M)$ be the set of faults in $M$ that can be executed only with test drivers that are hard to identify, that is, \(F_{hard\_d}(M) \equiv \{f\in F(M) | \forall{t}: Exec(f, t) \Rightarrow Hard\_d(t)\}\).

Then  the controllability of module $M$ is quantified as
\[Contr(M) = 1 - \frac{|F_{hard\_d}(M)|}{|F(M)|}.\]
\end{definition}

\begin{definition}{\bf Idealistic Observability}\label{def:obs:ideal}
Let $F_{r}(M)\subseteq F$  be the set of faults that can be revealed with some test case, that is, \(F_{r}(M) \equiv \{f\in F(M) | \exists{t}: Reveal(f, t)\}\). Let  $Hard\_o: T \rightarrow \{true, false\}$ be a criterion to decide whether or not the oracle part of a test case is hard to be identified. The oracle is the part of a test case that evaluates the outputs against the specification to exclude or pinpoint malfunctions. Correspondingly, let  $F_{hard\_o}(M)$ be the set of faults in $M$ that can be revealed only with oracles that are hard to identify, that is, \(F_{hard\_o}(M) \equiv \{f\in F_{r}(M) | \forall{t}: Reveal(t, f) \Rightarrow Hard\_o(t)\}\). 
 
Then the observability of module $M$ is quantified as
\[Obs(M) = 1 - \frac{|F_{hard\_o}(M)}{|F_{r}(M)|}.\]
\end{definition}

As mentioned, these definitions capture an idealistic scenario, in which we know the potential faults and the possible test cases in advance, which is, of course, unrealistic. In the next subsection, we present a measurement framework that proxies these idealistic metrics by referring to concrete faults, concrete test cases and objective decisions on evaluating the hardness of test cases and faults.

\subsection{A Framework to Measure Evidence-Based Testability}
The measurement framework that we propose in this paper addresses testability of object oriented classes. It instantiates the idealistic testability metrics by exploiting
\begin{inparaenum}[(i)]
\item mutation-based fault seeding to proxy the potential faults in the classes, 
\item search-based test generation to heuristically sample the possible test cases, 
\item weak-kill analysis and 
\item de-encapsulation to discriminate hard-to-identify test drivers and test oracles.

\end{inparaenum}

\textit{Mutation-based fault seeding} instruments programs with possible faults by using mutation operators, each describing a class of code-level modifications that may simulate faults in the program~\cite{demillo1978hints,pezze2008software}.
For instance, \emph{replacing numeric literals} is a mutation operator that produces different program versions (called \emph{mutants}) by changing a numeric literal in the program with a compatible literal: It produces a mutant for each possible legal replacement.
A test case that has a different outcome when executed against either the original program or a mutant $m$ is said to \emph{kill the mutant $m$}, meaning that it reveals the sample fault that the mutant represents. 
Several researchers argue that  mutants are valid representative of real faults~\cite{andrews2005mutation}. 
Our measurement framework refers to the set of mutation operators defined in the tool Major~\cite{just2014major}, and exploits Major to both generate the mutants and analyze killed mutants.

We compute test cases based on the \textit{search-based test generator} EvoSuite~\cite{fraser2011evosuite}, which samples the possible test cases with meta-heuristic algorithms guided with fitness functions based on coverage criteria, and generates test cases with assertion-style oracles on the observed outputs. In our evaluation framework, EvoSuite plays the role of a reference tester that works with the same capability consistently. 

Mutation operators may generate a large amount of mutants, including mutants that we cannot execute due to the intrinsic limitations of the test generator of choice, rather than because of testability issues. Our framework takes a twofold approach to coping with this. On one hand, we focus only on mutants that can be executed with at least a generated test case, even if not necessarily revealed. Technically, we rely on the notion of \textit{weak-kill analysis} (that belongs to the theory of \textit{mutation analysis}) as provided in Major. A test case weakly-kills a mutant if there
exists at least an execution state that differs between the original program and the mutant.

On the other hand  we exploit \textit{de-encapsulation}, to empower the test generator to directly set any input and state variable of any class in the program, and thus produce input states that may be hard generate otherwise.  
De-encapsulation is achieved by augmenting the interface of the  classes under test with custom setters for all class variables. Next, we execute the test generator on both the original classes and their de-encapsulated versions, and determine the set of executable mutants as the ones weakly-killed with at least an obtained test case.
We are aware that, technically speaking, using de-encapsulated classes may lead us to generate some input states that are illegal for the original classes. Nonetheless, our measurement framework embraces this approach heuristically: observing faults that the test generator can execute only with de-encapsulation is a sign of strict class interfaces, which may pinpoint testability issues.  

We estimate the hard-to-execute and hard-to-reveal mutants (the sets $F_{hard\_o}$ and $F_{hard\_d}$ of Definitions~\ref{def:contr:ideal} and \ref{def:obs:ideal}) as the mutants that we could either execute only with custom setters or weakly-kill but not ultimately kill, respectively.
The rationale is that the mutants that could be executed only with custom setters might, at least in principle, be in the scope of the test generator, although it seems hard for our reference tester to identify test cases that execute those mutants under the testability constraints imposed by the actual interfaces.
Similarly, mutants that we detect as weakly-killed but not ultimately killed with any test case, provide evidence of faults that can be in principle revealed, but may require oracles that are hard for the test generator to identify.

In summary, let $\hat{F}_{kill}$, $\hat{F}_{wkill}$ and $\hat{F}_{wkill\_noset}$ be the set of mutants that the generated test cases kill, weakly kill and weakly kill even without using custom setters, respectively; 
our testability measurement framework makes the following estimates related to the sets in Definitions~\ref{def:contr:ideal} and \ref{def:obs:ideal}:

\begin{itemize}
\item 
$\hat{T}$, all test cases that EvoSuite generates in bounded time for the target classes with and without custom setters. 
\item $\hat{F} \equiv \hat{F}_{r} \equiv \hat{F}_{wkill}$, all mutants executed (at least weakly-killed) with at least a test case $t\in \hat{T}$.
\item $\hat{F}_{hard\_d} \equiv \hat{F}_{wkill} - \hat{F}_{wkill\_noset}$, the mutants executed (at least weakly-killed) only with test cases with custom setters.
\item $\hat{F}_{hard\_o} \equiv \hat{F}_{wkill} - \hat{F}_{kill}$, the mutants that can be weakly-killed, but not killed.
\end{itemize}

\medskip
We then estimate controllability and observability as:

\begin{definition}{\bf Estimated Controllability}
We estimate the controllability of a class $C$ as the portion of mutants executed only with test cases that do not rely on custom setters:
\[Contr(C) = 1 - \frac{|\hat{F}_{hard\_d}(C)|}{|\hat{F}(C)|} = \frac{|\hat{F}_{wkill\_noset}(C)|}{|\hat{F}_{wkill}(C)|}.\]
\end{definition}
\begin{definition}{\bf Estimated Observability}
We estimate the obsevability of a class $C$ as the portion of mutants that were weakly-killed, but not killed:
\[Obs(C) = 1 - \frac{|\hat{F}_{hard\_o}(C)|}{ |\hat{F}_{r}(C)|} 
= \frac{|\hat{F}_{kill}(C)|}{|\hat{F}_{wkill}(C)|}.\]
\end{definition}

At the current state of our research, we do not yet provide an estimation for the overall testability  (Definition~\ref{def:testability}) of a class, since our conservative assessments of the sets of executable faults (for which we allow test cases with custom setters) and revealed faults (for which we accept weak kill as sufficient evidence), respectively, do not match well the precise way in which the controllability and observability facts combine into testability in the ideal scenario of Definition~\ref{def:testability}. 

\subsection{Prototype Implementation}
We have built a testability measurement prototype on top of the mutation analysis tool Major~\cite{just2014major}, and the test generator EvoSuite~\cite{fraser2011evosuite}.
We exploited the \textit{JavaParser} code-manipulation library~\cite{smith2017javaparser} to generate classes augmented with custom setters.

Our prototype executes EvoSuite six times for the original classes and six times on the classes with the custom setters; therefore,  twelve times in total. For each six-run group, it executes EvoSuite twice for each of three fitness functions, aiming to improve code coverage and mitigate the impact of the random choices of the search-based algorithm of EvoSuite. Namely we refer to the following fitness functions: 
\begin{inparaenum}[(i)] 
\item line and branch coverage,
\item weak mutation coverage, and
\item the composition of these two.
\end{inparaenum}
We set a time budget of 10 minutes for each run of EvoSuite, and relied on the functionality of the tool to generate test cases that include assertions on the observed outputs.\footnote{We used EvoSuite with the option \emph{assertion\_strategy=ALL.}}
For each class, we consider the union of all test cases that EvoSuite generated in the twelve runs.

Next, our prototype executes Major to collect the statistics on the killed and weakly-killed mutants by executing the test cases in two passes, in which we retain or strip off the custom setters, respectively. We ignore the mutants that  belong to the code of the custom setters.

We remark that, even if the current prototype is arguably slow, in building this first implementation we gave priority to being able to gather initial data on whether our measurements can reflect actual testability issues. We leave for future work the challenge of devising an efficient implementation.

%% file: section/exp.tex
\textbf{Subjects:} We executed our prototype to evaluate the testability of Closure Compiler (commit 46897c4), a JavaScript compiler written in Java, which is part of the experimental benchmarks of Defect4J~\cite{just2014defects4j}. 
The codebase of Closure Compiler consists of 483 classes\footnote{We do not include the inner classes in this count.} comprised of 93,907 lines of code.
Our prototype generated a total of 51,707 mutants for 355 classes (while for 128 classes the mutation tool did not yield any mutant) and a total of 2,898 test cases.
We were able to classify 21,920 and 29,441 killed and weakly-killed mutants overall, respectively, and 28,847 mutants weakly-killed without using custom setters.
We executed our prototype on a cloud facility with a virtual machine equipped with Linux Ubuntu, 48 cores Intel Xeon 2.4 GHz, and 160 GB of RAM. The experiment took a total of 200 hours.

\textbf{Results} 
Our measurements indicated that 85\% of the classes of Closure Compiler have max controllability, while they distribute more evenly in the range of observability values, with less than 15\% of the classes scoring a max observability.
However, at the current state of our research, it is hard for us 
to quantitatively evaluate the precision of our measurements, because we miss a reference ground-truth.
We thus conducted a qualitative study on the classes with the lowest and highest $Contr$ and $Obs$ scores, to try to confirm our hypothesis that  low and high scores  according to our metrics can be reconciled to actual testability issues and  boosters, which we could reveal with manual inspection of the classes. 

The top part of Table \ref{table:controllability} and Table \ref{table:observability} summarize our findings for the 10 classes with the lowest controllability and  observability, respectively. The bottom part of the  tables report the 5 classes with the highest scores (the ones with most weakly-killed mutants among those). We considered only classes for which our prototype identified at least 10 weakly-killed mutants. 
The first four columns of the tables indicate the class name (column \emph{Class}), the lines of code  in the class (\emph{LOC}), the amount of weakly-killed mutants (\emph{\#wkill}), 
and the corresponding $Contr$ (resp. $Obs$) score, which is a portion of the weakly-killed mutants.
The last column reports our findings on the controllability (resp. observability) issues or boosters that we identified manually in the classes.

\begin{table}[t]
\footnotesize
\begin{center}
 \caption{Qualitative study of classes with low and high $Contr$ scores}
 \label{table:controllability}
 \begin{tabular}{l|ccl|l|}
 \bf Class & \bf LOC & \bf \#wkill &  \bf Contr & \bf Issues\\
 \hline
 UnreachableCodeElimination  & 146 & 96 & 0.01 & \bf (i1)\\
 FindExportableNodes  & 96 & 25 & 0.20 & n.a.\\
 SourceMapConsumerV2 & 98 & 95  & 0.30 & \bf(i1) (i2)\\
 PeepholeOptimizationsPass & 71 & 48  & 0.36 & \bf(i1)\\
 ObjectPropertyStringPostprocess & 46 & 37  & 0.47 &\bf (i3)\\
 VarCheck & 199 & 97  & 0.60 & \bf(i2)\\
 CheckGlobalNames & 145 & 32  & 0.62 & n.a.\\
 OptimizeArgumentsArray & 121 & 22 & 0.64 & \bf(i1)\\
 AnalyzeNameReferences & 84 & 25 & 0.80 & \bf(i1)\\ 
 FunctionTypeBuilder & 605 & 283 & 0.85 & \bf(i2)\\
 \hline
 \hline
 RegExpTree  & 1403 & 1356  & 1.00 & \bf (b1)\\
 Fuzzer & 739  & 864 & 1.00 & \bf (b1) (b2) \\
 JsMessage & 426 & 656  & 1.00 & \bf (b1) (b2)\\
 CharRanges & 311 & 521  & 1.00 & \bf (b1) (b2)\\
 SourceMapGeneratorV2 & 421 & 458  & 1.00 & \bf (b1)\\ 
  \hline
 \end{tabular}
 \end{center}

\begin{tabular}{p{.96\columnwidth}}
\textbf{ISSUES}\\
\textbf{(i1) Multi-step protocol}: The class interface induces an interaction protocol that requires test cases to call multiple methods in specific sequences to set the relevant input states, thus hardening the task of identifying  tests.\\  

\textbf{(i2) Complex structured inputs}: The class takes several inputs defined as complex data structures, and thus requires long  test cases that go through   sophisticated initialization sequences to set the relevant inputs.\\  

\textbf{(i3) Preconditioned updates}: The interface methods  for updating the class state are guarded with many preconditions, and thus the class challenges the testers to comply with the preconditions when specifying the test inputs. \\ 
\textbf{(n.a.)} We did not identify any controllability issue.\\

~\\
\textbf{BOOSTERS}\\
\textbf{(b1) Simply-typed control inputs}: The class methods are fully controllable with inputs of primitive type, string type or types defined as flat data structures with only primitive fields and setters for all fields. \\
\textbf{(b2) Complete field-input mapping in constructors}: The test cases can rely on class constructors to assign all fields based on simply-typed inputs of the constructors, one input for each field. \\
\end{tabular}
\end{table}

\begin{table}[t]
\footnotesize
\begin{center}
 \caption{Qualitative study of classes with low and high $Obs$ scores}
 \label{table:observability}
 \begin{tabular}{l|ccl|l|}
 \bf Class &  \bf LOC & \bf \#wkill  & \bf Obs & \bf Issues\\
 \hline
BooleanType & 49 & 11  & 0.00 & \bf(i1)\\
SourcePosition & 23 & 17 & 0.06 & \bf(i2)\\
NullType & 53 & 12 & 0.08 & \bf(i1)\\
FunctionBuilder & 82 & 22 & 0.09 & n.a.\\
JvmMetrics & 214 & 81 & 0.16 & \bf(i2)\\
CommandLineRunner & 405 & 37 & 0.16 & \bf(i3)\\
Timer & 55 & 17 & 0.18 & \bf(i2)\\
XtbMessageBundle & 138 & 10 & 0.20 & \bf(i4)\\
Base64VLQ & 54 & 29 & 0.24 & n.a.\\
GraphPruner & 53 & 18 & 0.28 & \bf(i1) \bf(i3)\\
 \hline
 \hline
Token & 212 & 101 & 1.00 & \bf (b1)\\
NameReferenceGraphReport & 145 & 77 & 1.00 & \bf(b1) (b2) \\
BooleanLiteralSet & 39 & 14 & 1.00 & \bf (b1) \\
SimpleDependencyInfo & 42 & 13 & 1.00 & \bf (b1) (b2)\\
DiagnosticGroupWarningsGuard & 29 & 11 & 1.00 & \bf (b1)\\
\hline
 \end{tabular}
 \end{center}

\begin{tabular}{p{.96\columnwidth}}
    \textbf{ISSUES}\\
    \textbf{(i1) Complex observer methods}: The observer methods depend on many parameters, or take complex data structures as input, thus hardening the task of specifying the test oracles.  \\
    \textbf{(i2) Output on system streams}: The class produces most output on system streams, e.g., it writes results to the console or in the GUI, hardening the task of specifying automatic test oracles on those results.\\
    \textbf{(i3) Structural updates on private data}: The class computes structural characteristics of internal data structures with private visibility, e.g., the dimension of internal arrays, which cannot be checked with test oracles.\\
    \textbf{(i4) Asynchronous updates}: The class delegates asynchronous methods, which then produce results with callbacks, but it is difficult for the test oracles to predicate on those results\\
\textbf{(n.a.)} We did not identify any observability issue.\\
~\\
\textbf{BOOSTERS}\\
\textbf{(b1) Output as simply-typed return values}: The class produces all relevant outputs as return values of simple types, which can be easily checked with test oracles. \\
\textbf{(b2) Full getter access to modified fields}: The class stores its outputs in fields that the test cases can easily access with provided getter methods.\\
 \end{tabular}
\end{table}

As expected, our metrics scored maximum values for classes that allow for controlling the execution with simple inputs of interface methods and constructors, and observing the results in return values or with getter methods. Conversely, controllability and observability issues depends on interfaces that hamper test cases from setting relevant input values and inspecting relevant outputs. We describe the specific issues and boosters in detail at the bottom of each table.
For two low $Contr$ classes, and two low $Obs$ classes, we marked the result as \textit{false positive}. We traced these low scores to  random behaviors of EvoSuite that missed easy-to-spot test inputs and oracles.
However, we indeed reconciled the $Contr$ and $Obs$ scores to actual controllability and observability issues and boosters for all other classes.
We observe that in both tables  the amounts of lines of code oscillate almost within the same range of values across the classes with either low or high scores, suggesting no clear relation between class size and class testability.

%% file: section/conclusion.tex
In this papers we discussed a new approach for measuring testability characteristics of object oriented classes. Our approach tackles the testability measurement problem in direct fashion, by sampling the fault space of the classes and discriminating the relative portions of faults that are either easy or hard to be executed or revealed with automatically generated test cases. Thus, our approach 
differs from previous work that attempted to measure testability based on unproven relations to test size, test coverage and fault sensitivity. We are currently working to devise an efficient implementation of our prototype of the measurement framework, and to design experiments for evaluating our metrics quantitatively. 